\documentclass{jaa}
\usepackage{natbib}
\usepackage{booktabs}
\usepackage{txfonts}
\usepackage{graphicx}
\usepackage{textcomp}
\usepackage{gensymb}
\usepackage{xcolor}
\usepackage{siunitx}
\usepackage[]{hyperref}
\hypersetup{
    colorlinks,
    citecolor=blue,
    filecolor=blue,
    linkcolor=blue,
    urlcolor=blue
}
\usepackage{multirow}

\newcommand{\pjfirst}[1]{#1}
\newcommand{\pjnew}[1]{#1}
\newcommand{\pjmore}[1]{#1}
\newcommand{\pjmoreii}[1]{#1}
\newcommand{\pjnewii}[1]{#1}

\newcommand{\pjfinal}[1]{#1}

\begin{document}\sloppy

\title{UVIT Data Release version 7: Regenerated high-level UVIT data products}


\author{Prajwel Joseph\textsuperscript{1,2,*}, S. N. Tandon\textsuperscript{3}, S. K. Ghosh\textsuperscript{4} and C. S. Stalin\textsuperscript{1}}
\affilOne{\textsuperscript{1}Indian Institute of Astrophysics, Bangalore 560034, India\\}
\affilTwo{\textsuperscript{2}Department of Physics and Electronics, CHRIST (Deemed to be University), Bangalore 560029, India\\}
\affilThree{\textsuperscript{3}Inter-University Centre for Astronomy and Astrophysics, Pune 411007, India\\}
\affilFour{\textsuperscript{4}Tata Institute of Fundamental Research, Mumbai 400005, India\\}


\twocolumn[{

\maketitle

\corres{prajwelpj@gmail.com}

\msinfo{19 August 2024}{1 January 2025}

\begin{abstract}
The Ultra-Violet Imaging Telescope (UVIT) on board \textit{AstroSat} is \pjfirst{an} active telescope capable of high-resolution far-ultraviolet imaging ($<$\ang{;;1.5}) and low resolution ($\lambda/\delta\lambda$ $\approx$ 100) slitless spectroscopy with a field of view as large as $\sim$0.5 degrees. 
Now almost a decade old, UVIT continues to be operational and generates valuable data for the scientific community. 
UVIT is also capable of near-ultraviolet imaging ($<$\ang{;;1.5}); however, the near-ultraviolet channel stopped working in August 2018 after providing data for nearly three years.
This article gives an overview of the latest version (7.0.1) of the UVIT pipeline and \pjfinal{UVIT Data Release version 7.}
The high-level products generated using pipeline versions having a major version number of seven will be called ``\pjfinal{UVIT Data Release version 7}". 
The latest pipeline version overcomes the two limitations of the previous version (6.3), namely (a) the inability to combine all episode-wise images and (b) the failure of the astrometry module in a large fraction of the observations.
The procedures adopted to overcome these two limitations, as well as a comparison of the performance of this new version over the previous one, are presented in this paper.
\pjfinal{The UVIT Data Release version 7} products are available at the Indian Space Science Data Center of the Indian Space Research Organisation for archival and dissemination from 01 June 2024.
The new pipeline version is open source and made available on GitHub.
\end{abstract}


}]


\doinum{12.3456/s78910-011-012-3}
\artcitid{\#\#\#\#}
\volnum{000}
\year{0000}
\pgrange{1--}
\setcounter{page}{1}
\lp{1}

\section{Introduction}
The Ultra-Violet Imaging Telescope (UVIT, \citealt{tandon2017orbit}) onboard \textit{AstroSat}, the first Indian multi-wavelength space observatory, started observing the Universe almost a decade ago \citep{singh2014astrosat}.
UVIT continues to provide ultraviolet (UV) imaging and low-resolution spectroscopy data of many sky regions, driven by the proposals submitted through the \textit{AstroSat} Proposal Processing System (APPS, \citealt{balamurugan2021astrosat}). 
UVIT can observe simultaneously in the far-UV (FUV, 1300--1800 \AA) and near-UV (NUV, 2000--3000 \AA) channels using photon counting detectors \pjfirst{with a field of view as large as $\sim$0.5 degrees} \citep{kumar2012ultraviolet, postma2011calibration}. 
Images are also acquired in the visible (VIS, 3200--5500 \AA) channel every second for correcting the pointing drift for each individual episode of imaging. 
The point sources in most drift-corrected UVIT UV images have a full-width half maximum (FWHM) of $<$\ang{;;1.5} \pjnew{\citep{tandon2017inorbit, tandon2020additional}}.
\pjfirst{Each channel is equipped with filters that allow the selection of specific wavelength bands within that channel.}
In UV channels, different partial-field window modes are available apart from the full-field mode, where smaller window modes enable faster sampling of images of the selected window\footnote{Please see page 26, \textit{AstroSat} handbook (available at \url{http://www.iucaa.in/~astrosat/AstroSat_handbook.pdf}).}.  

The UVIT payload operations centre (POC) at the Indian Institute of Astrophysics (IIA), together with the various centres of the Indian Space Research Organisation (ISRO), carries out all operations related to UVIT. 
The POC generates high-level UVIT data products (Level2 products) using the UVIT Level2 pipeline \citep{ghosh2022automated} from the low-level UVIT (Level1) data.
The Level1 products mainly consist of UV photon counting data and $\sim$1-second VIS images without pointing drift correction.
The Level2 data contains UV photon counting data with flat-field, distortion and drift corrections, along with other derived products such as count-rate and error images and exposure maps.
UVIT Level2 products are available through the Indian Space Science Data Center (ISSDC) \textit{AstroSat} Archive \citep{balamurugan2021data}.
All \textit{AstroSat} data becomes public after a stipulated proprietary period\footnote{Please see \url{https://www.issdc.gov.in/docs/as1/astrosat-data-guidelines.pdf}}.
The POC monitors the instrument's health, and instrument problems are addressed with the help of ISRO.
Except for the loss of the NUV channel in 2018 and other occasional problems \pjfirst{(see \citealt{ghosh2021orbit} for a list of known problems encountered by UVIT upto 2021), the FUV and VIS channels are working normally as of 2024.}
When possible, the problem-affected data is fixed on the ground using purpose-built software in the POC.
Additionally, the POC develops and maintains tools for feasibility checking of the UVIT observation proposals (CanUVIT, \citealt{canuvit}), Level1 data inspection and correction (UVITility, \citealt{uvitility}), and Level2 data analysis (Curvit, \citealt{joseph2021curvit}; UVIT Level2 Pipeline, \citealt{ghosh2022automated}). 

\pjfirst{Among the UV missions with photon-counting detectors (e.g., Hubble Space Telescope Cosmic Origins Spectrograph \citep{green2011cosmic}; \textit{Swift} Ultra-Violet/Optical Telescope \citep{roming2005swift}; XMM-Newton Optical/UV Monitor \citep{mason2001xmm}), UVIT is most comparable to the Galaxy Evolution Explorer (GALEX, \citealt{martin2005galaxy}) in terms of capabilities. 
GALEX operated in the FUV (1344--1786 \AA) and NUV (1771--2831 \AA) bands, featuring a \ang{1.25} field of view and a resolution of $\sim$4--\ang{;;6} \pjfinal{\citep{morrissey2005orbit}}.
During its mission, GALEX conducted near-complete sky surveys in both bands \citep{bianchi2014galaxy}.
Photon-counting data from GALEX was processed with a dedicated pipeline to generate high-level products such as count-rate images and exposure maps (for details on the GALEX data processing and pipeline products, see \citealt{morrissey2007calibration}).
The extensive use of the GALEX data owes greatly to the accessibility of these high-level, science-ready products, which allowed researchers to focus on their science cases without delving into the complexities of data reduction.
Similarly, providing high-level products for UVIT should significantly enhance the usability and adoption of UVIT data by the astronomical community.
Presently, UVIT data users either directly use the high-level products available from the ISSDC \textit{AstroSat} Archive\footnote{\url{https://astrobrowse.issdc.gov.in/astro_archive/archive/Home.jsp}} or reduce the UVIT Level1 data themselves using the various UVIT pipelines available \citep{postma2017ccdlab, murthy2017jude, ghosh2022automated}.
}

The previous 6.3 version of the UVIT Level2 pipeline (hereafter UP6) has been used by the POC to generate Level2 products from 2019 to 2024.
It has addressed or worked around various problems in the Level1 data \citep{ghosh2021performance}.
However, there are two major limitations to UP6. 
When observations are acquired in multiple orbits, all the orbit-wise Level2 photon counting data must be combined to generate combined data products.
This combining process sometimes fails.
The second limitation is the frequent failure of the UP6 astrometry module to find astrometric solutions.
There were minor problems, too, such as the patterns in images that originate from rotation and the incorrect treatment of the exposure map edge regions during the combining process. 
UVIT Level2 pipeline version 7.0.1\footnote{The POC has adopted a MAJOR.MINOR.PATCH versioning scheme compared to the previously used MAJOR.MINOR scheme.}(hereafter UP7) was developed by the POC to improve over these shortcomings. 
The POC is regenerating Level2 products for all observation IDs with available Level1 data ($>$1600 observation IDs till June, 2024) using UP7. 
The UVIT Level2 products generated using pipeline versions having a major version number of seven will be called ``\pjfinal{UVIT Data Release version 7}". 
\pjfinal{The UVIT Data Release version 7} products are available for all new observations processed at the POC starting from June 1, 2024.
\pjfinal{UVIT Data Release version 7} products will also be made available for observations taken prior to June 1, 2024, in a phased manner.
This article gives a brief overview of UP7 and \pjfinal{the UVIT Data Release version 7} products.

This paper is organised as follows. Section 2 gives an overview of UP7 and its major steps. 
A brief description of \pjfinal{the UVIT Data Release version 7} products are given in Section 3 followed by the Summary in the final Section.

\begin{figure*}
\centering
        \includegraphics[width=\textwidth]{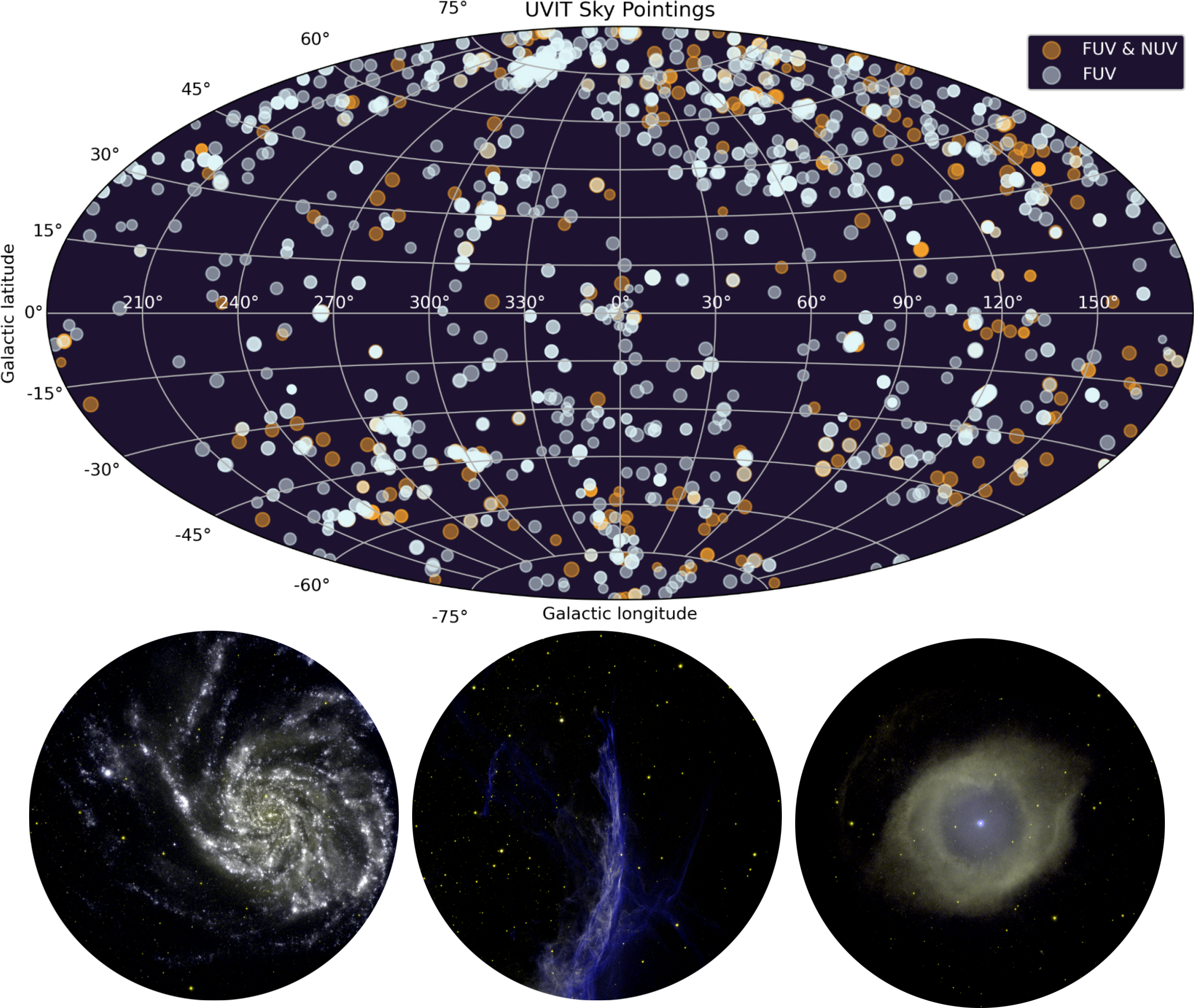}
    \caption{The top panel shows UVIT sky pointings till May 24, 2024. The observations with both FUV and NUV data are shown in orange, and observations with only the FUV data are shown in light blue.  
    \pjfinal{The circle radii are logarithmically proportional to the observation exposure times, ranging from a maximum of 206860s to a minimum of 120s.
    The filled circles are semi-transparent, making regions with multiple pointings appear more prominently coloured than those with a single pointing.}
    The bottom panel shows three pseudo-colour images of fields observed with UVIT; the M101 galaxy \pjfirst{(observation ID: G05\_233T10\_9000000570)} is shown on the left, A section of the Veil nebula \pjfirst{(observation ID: G05\_215T01\_9000000586)} in the middle, and the Helix nebula \pjfirst{(observation ID: G05\_187T01\_9000000690)} on the right. 
    In all images, the FUV channel is blue, and the NUV channel is yellow.
    \pjnew{The M101 galaxy image was created using F148W (exposure time: 3281s) and N219M (8097s) filter observations. 
    The Veil Nebula image combines observations with the F154W (3231s) and N279N (10633s) filters. 
    For the Helix Nebula, the FUV channel image used integrates F172M (1099s) and F169M (293s) filter observations, while the NUV image combines N245M (1108s) and N263M (302s) filters.
    }
    The images have been processed to bring out faint features.
             }
    \label{fig:mainfig}
\end{figure*}

\section{UVIT Level2 pipeline version 7.0.1}
\label{up7}

UP6 has been used to process over 1600 Level1 datasets.
It can automatically run on Level1 datasets to generate Level2 products \citep{ghosh2021performance, ghosh2022automated}. 
Fig. \ref{fig:mainfig} shows the UVIT sky pointings\footnote{The UVIT sky pointing data for each observation is taken from the \textit{AstroSat} Schedule Viewer website at \url{https://webapps.issdc.gov.in/MCAP/}. The data thus collected can be accessed at \url{https://github.com/prajwel/Arivu/blob/main/UVIT_pointings_and_filterwise_exptimes.csv}.} and three UVIT images created from the Level2 products. 
\pjfirst{The low density of UVIT pointings in the galactic plane compared to the poles is due to the presence of very bright stars in the galactic plane.
Observations are permitted only for those fields \pjnew{which do not have any sources beyond the acceptable brightness. For details on implementing this  constraint, please see the ``UVIT filter safety check" document\footnote{\url{https://www.iiap.res.in/documents/773/Mandatory\_checks\_to\_be\_done\_for\_UVIT\_observations.pdf}}.}}




The UVIT Level2 pipeline development was initially started in the C++ programming language.
The development in C++ continued till UP6.
In UP7\footnote{The pipeline can be accessed at \url{https://github.com/prajwel/UVPipe}.}, we have retained all of the UP6 C++ codes and developed additional features in Python.
The only modification to the UP6 C++ code used in UP7 is fixing a bug that caused inaccurate flat-field correction and bad event flagging in certain conditions \pjnew{(see \ref{sec:bug})}.
\pjnewii{For the bug-affected images, the erroneous flat-field corrections result in count-rate uncertainties of less than 1\% for sources located within a radius of 1900 sub-pixels from the centre in most NUV filter images, excluding the N219M and N279N filters.
For the N279N filter and all FUV filters, uncertainties remain below 5\%, whereas the N219M filter shows a maximum uncertainty of less than 10\%.}

A brief description of the various major steps of UP7 data processing of Level1 datasets is given below. 

\subsection{Pointing drift correction in episodes}
\textit{AstroSat} is in a low-earth orbit that takes around $\sim$90 minutes to complete.  
UVIT observes only during the dark phase of the orbit when sun is eclipsed by the earth, also bound by other constraints \citep{singh2021astrosat}. 
\pjfirst{In a scenario where the entire eclipse duration of each orbit is fully utilised for UVIT observations, with a single filter configured across all channels, the Level1 dataset is organised as follows:
\begin{itemize}
    \item The data is grouped by channel, with each channel represented as a parent directory.
    \item Within each channel’s directory, there are subdirectories corresponding to individual orbits.
    \item Each orbit-specific subdirectory contains the data acquired during that particular orbit.
\end{itemize}
}

However, there may be data splits (due to filter change or other reasons) in most cases, and data acquired in one orbit will be found across multiple directories.
 We refer to science data found in a single bottom-level Level1 directory as an episode for practical purposes.
 
\begin{figure}
\centering
        \includegraphics[width=\columnwidth]{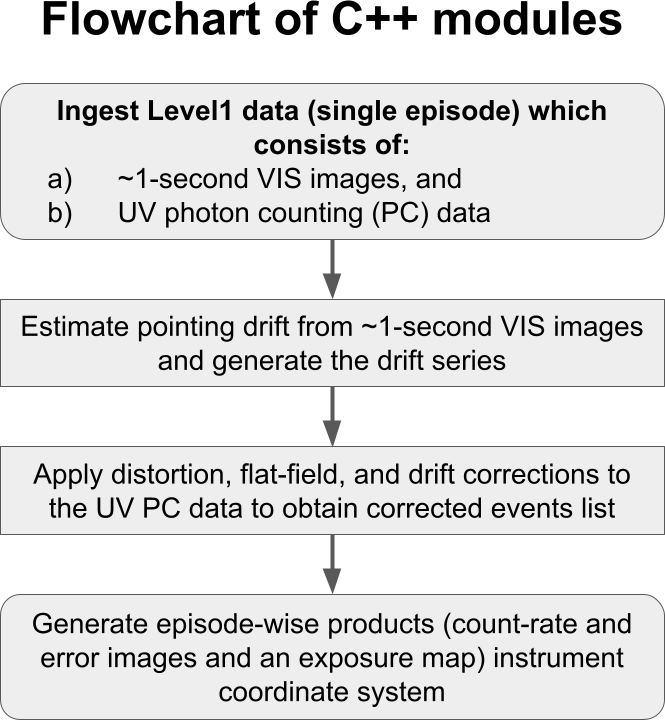}
    \caption{A simplified depiction of the scope of C++ modules in UP7 \pjfinal{as a flowchart}.}
    \label{fig:c_scope}
\end{figure}

\begin{figure}
\centering
        \includegraphics[width=\columnwidth]{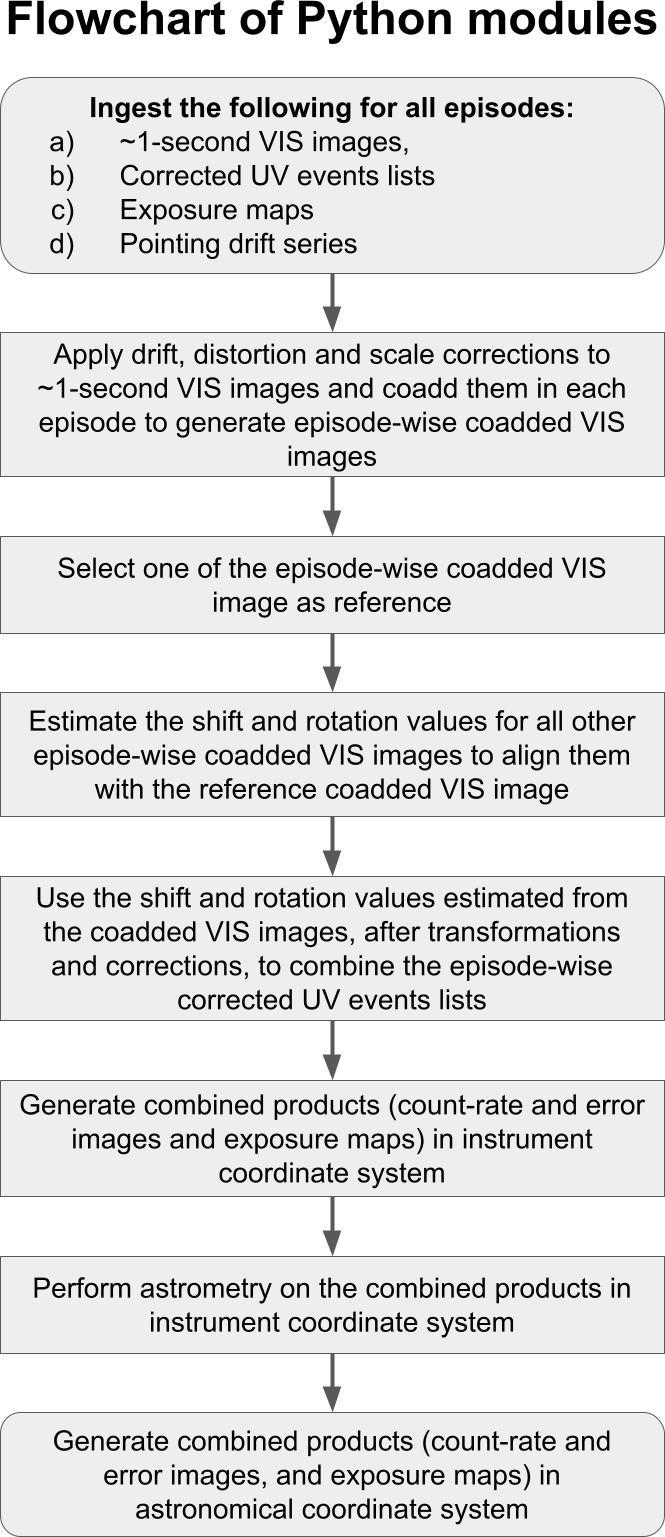}
    \caption{A simplified depiction of the scope of Python modules in UP7 \pjfinal{as a flowchart}.}
    \label{fig:python_scope}
\end{figure}

The UV data in each Level1 episode consists of time-tagged photon positions (events). 
In the full-field  window mode, events are detected using a centroiding algorithm from each detector readout frame of size 512$\times$512 pixels at $\sim$29 Hz \citep{hutchings2007photon}.
This centroiding algorithm provides sub-pixel resolution (1/8th of a detector pixel) for the event positions.
\pjfirst{UVIT UV images are generated with pixel sizes corresponding to this sub-pixel resolution.
Consequently, while the UV channel detectors have a native size of 512$\times$512 pixels, the final UV images are produced with a resolution of 4800$\times$4800 sub-pixels (8 $\times$ 512 = 4096 pixels plus an additional \pjnew{8 $\times$ 88 =} 704 pixels for padding\footnote{\pjnew{Here, padding refers to the addition of 44 pixels on either side of each axis of the original UV/VIS frame data in 512 $\times$ 512 pixels to account for the pointing drift. This process changes the data in 512 $\times$ 512 pixels to 600 $\times$ 600 pixels.}} to accommodate pointing drift along each axis).
The term ``sub-pixels" is used in UVIT-related literature to distinguish the UVIT UV image pixels from the original detector pixels. 
The UV channel plate scales are given in Table \ref{tab:my-table}.}
Due to the drift in telescope pointing, the position of any source in the sky drifts on the detector.
UP7 estimates the drift of each UV frame using the VIS images taken every $\sim$1 second relative to a chosen reference frame and corrects this in the coordinates of each event of the frame.
\pjfirst{The UV channels operate at a frame rate of $\sim$29 Hz or higher, depending on the user-selected window mode. The pointing drift is assumed to vary linearly within each $\sim$1-second \pjfinal{interval} of the VIS channel derived drift series. The drift series is linearly interpolated during the drift correction of UV events, aligning the event positions from different frames within an episode to a reference frame.}
The drift estimation from VIS data, correction of UV events and the generation of exposure maps are done by the UP7 incorporated C++ code of UP6 as described in \cite{ghosh2022automated}. 
\pjfirst{A simplified depiction of the scope of C++ modules in UP7 is shown as a flowchart in Fig. \ref{fig:c_scope}}. 
\pjfirst{All remaining steps, from subsection \ref{subsec:combining} to the generation of the \pjfinal{UVIT Data Release version 7} products, use the newly developed Python scripts part of UP7 and are shown as a flowchart in Fig. \ref{fig:python_scope}}.

\subsection{Combining episode data}
\label{subsec:combining}
The drift-corrected UV episode data are combined for each channel-filter-window combination in a particular observation.
For this purpose, the \pjfirst{shifts ($\delta_{uv,N}$) and rotations ($\rho_{uv,N}$)} between reference frames of \pjfirst{N number of UV episodes with a common channel-filter-window configuration} are estimated through the process given below:
\begin{enumerate}
    \item \textit{Generate coadded VIS images:} The term ``coadd" refers to the process of adding together drift-corrected $\sim$1-second VIS images in each episode.
    \pjfirst{The drift correction of these $\sim$1-second VIS images uses the drift series generated in the previous step (see \citealt{ghosh2022automated} for details).}
    Please note that the $\sim$1-second VIS images in 512$\times$512 pixels are scaled to 1800$\times$1800 pixels \pjnew{(3 $\times$ 512 = 1536 pixels plus an additional 3 $\times$ 88 = 264 pixels for padding to accommodate pointing drift along each axis)} and \pjfirst{corrected for drift and distortion} before coadding.
    \pjfirst{The $\sim$1-second VIS images are rescaled \pjfinal{to 1800$\times$1800 pixels from 512$\times$512 pixels} to improve accuracy of the multiple corrections applied.}  
    Also, a correction is applied to ensure almost equal plate scales along the X and Y axes.
    \pjfirst{The corrections are applied to the $\sim$1-second VIS images by converting the images to position-value list, applying the corrrections to the positions, and then converting back the modified position-value list to images.
    The corrected $\sim$1-second VIS images are coadded in each episode to generate coadded VIS images for every episode.}     
    The final coadded VIS images are in 1800$\times$1800 pixels.

    \item \textit{Estimate shift and rotation from coadded VIS images:} The \pjfirst{shifts ($\delta_{vis,N}$) and rotations ($\rho_{vis,N}$)} between the coadded VIS images are estimated with one of the coadded images as the reference episode. 
    The sources are detected in the coadded VIS images employing the DAOFIND algorithm \citep{stetson1987daophot}, and then \pjfirst{$\delta_{vis,N}$ and $\rho_{vis,N}$} are estimated with the Aafitrans Python package \citep{beroiz2020astroalign, aafitrans} using the detected sources as input. 
    All coadded VIS images are sorted based on the corresponding UV episode exposure times.
    The reference coadded VIS image will correspond to the UV episode with the maximum exposure time.
    \item \textit{Improve shifts using the correlation technique:} 
    Ideally, the \pjfirst{$\delta_{vis,N}$ and $\rho_{vis,N}$} estimated from coadded VIS images can be directly applied to the UV channels after transformations to account for the VIS and UV channel orientation differences.
    Transformation of the VIS channel to the UV channel orientation is an affine transformation that involves shift ($\delta_A$), rotation ($\rho_A$), scale ($S_A$), and reflection ($Ref_A$). 
    \pjfirst{This transformation can be represented as:
    \begin{align}
        \delta_{uv, N, ideal} &= \text{affine}(\delta_{vis,N},\ \delta_A,\ \rho_A,\ S_A,\ Ref_A),  \\
        \rho_{uv, N, ideal} &= \text{affine}(\rho_{vis, N},\ \delta_A,\ \rho_A,\ S_A,\ Ref_A).        
    \end{align}}
    
    While we had originally assumed that ($\delta_A$, $\rho_A$, $S_A$, $Ref_A$) would remain constant, our analysis revealed that $\delta_A$ for the VIS-to-FUV transformation significantly varies between episodes. 
    \pjnew{The variation in the VIS-to-FUV transformation is likely due to the cumulative effects of thermal stick-slip \citep{postma2021uvit}.}
    \pjfirst{The variation in $\delta_A$ means that we cannot directly use $\delta_{uv, N, ideal}$ and $\rho_{uv, N, ideal}$.
    To account for the $\delta_A$ variation, we assume that:
    \begin{align}
        \delta_{uv, N} &= \delta_{uv, N, ideal} + \delta_{uv, N, residual},  \\
        \rho_{uv, N} &= \rho_{vis, N}.        
    \end{align}}
    \pjfirst{We estimate $\delta_{uv, N, residual}$ using a correlation technique described below: 
    \begin{itemize}
        \item Drift-corrected UV events ($UV_{events, N}$) from N episodes are taken as input. 
        \item The positions of $UV_{events, N}$ are transformed by applying $\delta_{uv, N, ideal}$ and $\rho_{uv, N}$ to obtain $UV^{'}_{events, N}$.
        \item $UV^{'}_{events, N}$ are collapsed along the X and Y axes and correlated across episodes to obtain $\delta_{uv, N, residual}$ values. 
    \end{itemize}
    }
    
    Three different methods employing the correlation technique exist. 
    The method used for combining UV episodes can be found in the combined Level2 product FITS headers as a string value to the \texttt{COMBMETH} key. The key can have the following values. 
    \begin{enumerate}
        \item \texttt{DEFAULT\_METHOD}: The method uses all events data for the correlation technique. This is the default method.  
        \item \texttt{ALT\_METHOD1}: The first alternate method. 
        This method removes the background events from the events list before applying the correlation technique. 
        This method is useful for faint UV fields. 
        \item \texttt{ALT\_METHOD2}: The second alternate method. 
        Only the data of a bright source is used for correlation.
        This method is useful for grating observations and fields with extended sources.         
    \end{enumerate}
\end{enumerate}

\begin{figure}
\centering
        \includegraphics[width=\columnwidth]{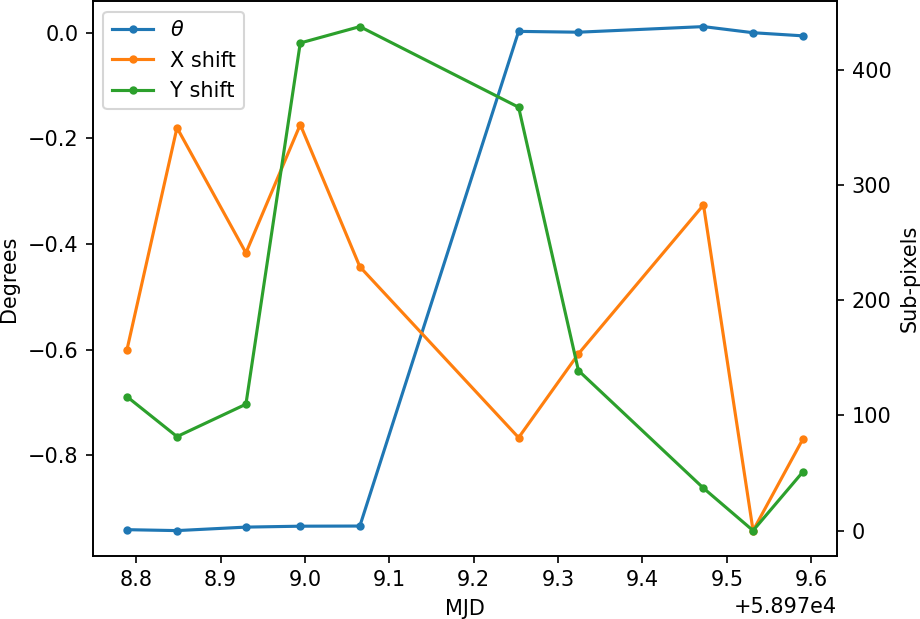}
    \caption{\pjnew{The shifts ($\delta_{uv,N}$) and rotations ($\rho_{uv,N}$) estimated for a set of episodes having a common channel-filter-window configuration.
    The $\delta_{uv,N}$ values are given as shifts along the X and Y axes, and $\rho_{uv,N}$ is given in degrees.
    \pjmore{The jump in $\rho_{uv,N}$ is due to the \textit{AstroSat} attitude correction manoeuvre \pjmoreii{to account for the $\sim$1-degree/day apparent motion of the sun} \citep{singh2021astrosat}.}
    }}
    \label{fig:shifts_rotations}
\end{figure}

Combined and time ordered events lists are generated for each channel-filter-window combination by combining all the episode-wise events lists using \pjfirst{the estimated $\delta_{uv,N}$ and $\rho_{uv,N}$ values to modify the event positions.
Fig. \ref{fig:shifts_rotations} shows the $\delta_{uv,N}$ and $\rho_{uv,N}$ values for a 
set of episodes having a common channel-filter-window configuration.
Combined exposure maps are also generated using the $\delta_{uv,N}$ and $\rho_{uv,N}$ values.}
Finally, count-rate and error images are generated from the combined events lists and exposure maps. 

In addition to the combined UV products, combined VIS images are generated by combining all the coadded VIS images. 
However, these are intended only for quick-look purposes and as alternate sources of astrometric solutions.

\subsection{Astrometry on combined products}
UP7 uses \texttt{Astrometry.net} software to obtain astrometric solutions for the combined products \citep{lang2010astrometry}.
Apart from the \pjfirst{5200-series index files derived from the Tycho-2 \citep{hog2000tycho} and Gaia data release 2 \citep{brown2018gaia} catalogues} provided by the \texttt{Astrometry.net} team, UP7 also uses the index files generated from a subset of the Gaia data release 3 catalogue for UVIT astrometry (\citealt{gaia2016, gaia2023}, Akanksha et al., in preparation). 
UP7 does not carry out astrometry on the episode-wise products.
While astrometric solutions are obtained for all combined products, there are differences between the UV and VIS astrometry processes.

In the VIS channel astrometry, \texttt{Astrometry.net} estimated astrometric solutions are directly used. 
The plate scale is allowed to vary\footnote{As far as the authors are aware, there is no option in the \texttt{Astrometry.net} software to obtain an astrometric solution using user-provided fixed plate scale values. However, it is possible to constrain the range of plate scale values.} in the astrometry solutions.
Corrections applied during the VIS coadding process ensure that the VIS plate scales are almost equal along the X and Y axes. 
Absolute plate scales have not been estimated for the VIS images.

The UV channel astrometry is a two-stage process.
First, \texttt{Astrometry.net} is used to obtain a list of UV image source positions and corresponding Gaia catalogue entries (image-to-catalogue list).
This list is then used to derive final astrometric solutions with known fixed plate scales of the UV images (see Table \ref{tab:my-table}).
If successful, this is noted in the image header with the \texttt{ASTMETH} keyword having the value \texttt{Direct}.
If unsuccessful, image-to-catalogue lists from the combined VIS images are transformed to the combined UV image orientation using known nominal VIS-UV alignments. 
Fixed plate scale astrometric solutions are then obtained for the UV images from the transformed list. 
When this indirect method is used, the \texttt{ASTMETH} keyword will have the value \texttt{Indirect}.
The \texttt{Indirect} astrometry solution for the combined UV images could have errors up to 5 arcseconds.
Using the astrometric solution thus obtained for the UV images, astrometric coordinates are provided for each event in the combined events list.
\pjfirst{The \texttt{Direct} astrometry method is generally successful, with failures primarily occurring in grating observations or images with very low exposure times.}

Originally, the UV images generated from the events lists were found to have different plate scales along the X and Y axes. 
To avoid differences in plate scales along the X and Y axes, UP7 modifies the photon positions in the events lists so that the generated UV images have a single plate scale along both axes. 
The UV image plate scales thus fixed are given in Table \ref{tab:my-table}.
The fixed plate scales were arrived at by finding the different plate scales along the X and Y axes for all filters in FUV and NUV channels and taking the average of X-scale and Y-scale for each filter.
Notice that all FUV filters have the same average plate scale.
However, this is not true for the NUV channel. 
The N242W filter was found to have a slightly different plate scale.
The N219M filter has a significantly different plate scale due to the filter having some optical power.
Also, the existing distortion corrections for the N219M filter are not as good as other filters.  
The relatively poor distortion corrections in the N219M filter lead to $>$2 times the root mean square (RMS) image-to-catalogue position differences in this filter than others. 


\begin{table}
\caption{UVIT UV channel plate scales. Please see Table 3 from \cite{joseph2021curvit} for information on the UVIT filter naming conventions.}
\label{tab:my-table}
\begin{tabular}{lr}
\hline
 & \multicolumn{1}{c}{Plate scale} \\ \hline
All FUV filters & \ang{;;0.41686}/sub-pixel \\
\begin{tabular}[c]{@{}l@{}}All NUV filters\\ (except N242W and N219M)\end{tabular} & \ang{;;0.41726}/sub-pixel \\
N242W & \ang{;;0.41733}/sub-pixel \\
N219M & \ang{;;0.41944}/sub-pixel \\ \hline
\end{tabular}%
\end{table}

\subsection{Pipeline Comparison}
To showcase the performance of the new 7.0.1 pipeline version, we compared the exposure times and astrometric accuracy of the combined UV images between the new and old pipeline versions.
The comparison was made for five observations, as shown in Table \ref{tab:performance}.
It can be seen from the comparison that the episode-combining efficiency and astrometric accuracy have significantly improved in the new version of products.

\begin{table*}
\caption{Comparison of the exposure times and astrometric accuracy of the combined UV images between the new and old pipeline versions. The V6.3 astrometric module failed on the FUV channel images for some observation IDs.
Therefore, two columns are shown to provide the astrometric errors in both cases separately (astrometry module successful and failed cases).
When the astrometric module fails in V6.3, the resulting combined images exhibit only a crude astrometric accuracy due to the astrometric information being taken from the spacecraft auxiliary data, with errors ranging from a few arcminutes to degrees.}
\label{tab:performance}
\resizebox{\textwidth}{!}{%
\begin{tabular}{|r|rr|rr|ccc|clc|}
\hline
\multicolumn{1}{|c|}{\multirow{2}{*}{Observation ID}} &
  \multicolumn{2}{c|}{\begin{tabular}[c]{@{}c@{}}FUV total\\ combined time\\ (s)\end{tabular}} &
  \multicolumn{2}{c|}{\begin{tabular}[c]{@{}c@{}}NUV total\\ combined time\\ (s)\end{tabular}} &
  \multicolumn{3}{c|}{\begin{tabular}[c]{@{}c@{}}FUV\\ astrometric\\ accuracy\end{tabular}} &
  \multicolumn{3}{c|}{\begin{tabular}[c]{@{}c@{}}NUV\\ astrometric\\ accuracy\end{tabular}} \\ \cline{2-11} 
\multicolumn{1}{|c|}{} &
  \multicolumn{1}{c|}{V6.3} &
  \multicolumn{1}{c|}{V7.0.1} &
  \multicolumn{1}{c|}{V6.3} &
  \multicolumn{1}{c|}{V7.0.1} &
  \multicolumn{1}{c|}{\begin{tabular}[c]{@{}c@{}}V6.3\\ (astrometry\\ module\\ successful)\end{tabular}} &
  \multicolumn{1}{c|}{\begin{tabular}[c]{@{}c@{}}V6.3\\ (astrometry\\ module\\ failed)\end{tabular}} &
  V7.0.1 &
  \multicolumn{2}{c|}{\begin{tabular}[c]{@{}c@{}}V6.3\\ (astrometry\\ module\\ successful)\end{tabular}} &
  V7.0.1 \\ \hline
C02\_008T01\_9000000884 &
  \multicolumn{1}{r|}{7862} &
  8108 &
  \multicolumn{1}{r|}{9738} &
  9738 &
  \multicolumn{1}{c|}{} &
  \multicolumn{1}{c|}{$\sim$\ang{;4}} &
  $<$\ang{;;1} &
  \multicolumn{2}{c|}{$\sim$\ang{;;4}} &
  $<$\ang{;;1} \\ \hline
G06\_134T03\_9000000894 &
  \multicolumn{1}{r|}{1384} &
  6772 &
  \multicolumn{1}{r|}{5382} &
  5841 &
  \multicolumn{1}{c|}{\ang{;;12}} &
  \multicolumn{1}{c|}{} &
  $<$\ang{;;1} &
  \multicolumn{2}{c|}{$\sim$\ang{;;2}} &
  $<$\ang{;;1} \\ \hline
G06\_164T01\_9000000896 &
  \multicolumn{1}{r|}{1702} &
  7673 &
  \multicolumn{1}{r|}{7733} &
  7733 &
  \multicolumn{1}{c|}{} &
  \multicolumn{1}{c|}{$\sim$\ang{;3}} &
  $<$\ang{;;1} &
  \multicolumn{2}{c|}{$<$\ang{;;1}} &
  $<$\ang{;;1} \\ \hline
A02\_082T01\_9000000898 &
  \multicolumn{1}{r|}{11749} &
  14744 &
  \multicolumn{1}{r|}{15760} &
  15760 &
  \multicolumn{1}{c|}{} &
  \multicolumn{1}{c|}{$\sim$\ang{1}} &
  $<$\ang{;;1} &
  \multicolumn{2}{c|}{$<$\ang{;;1}} &
  $<$\ang{;;1} \\ \hline
G06\_164T01\_9000000922 &
  \multicolumn{1}{r|}{1841} &
  7436 &
  \multicolumn{1}{r|}{7371} &
  7439 &
  \multicolumn{1}{c|}{} &
  \multicolumn{1}{c|}{$\sim$\ang{;3}} &
  $<$\ang{;;1} &
  \multicolumn{2}{c|}{$\sim$\ang{;;1}} &
  $<$\ang{;;1} \\ \hline
\end{tabular}%
}
\end{table*}

\section{UVIT Data Release version 7 products}
\label{arivu}
The POC has adopted UP7 from June 1, 2024 to process all new Level1 datasets to generate Level2 products.
The UVIT Level2 products generated using pipelines with a major version of 7 will be called ``\pjfinal{UVIT Data Release version 7}". 
\pjfirst{The new high-level data products represent a significant enhancement and reorganisation, incorporating several key improvements:
\begin{itemize}
 \item Introduction of new combined data products.
 \item \pjnew{Headers in the combined products now provide clear, concise, and user-friendly information.
 Key details have been segregated for better visibility, ensuring users can quickly access essential metadata.} 
 \item Improved organisation of the products for easier accessibility and quicker navigation.
 \item Updates and refinements to accompanying files such as \texttt{disclaimer.txt} and \texttt{README.txt}.
 \item Addition of quality metrics to enable users to assess the data effectively (described in \texttt{README.txt}).
\end{itemize}
Given the scale of these updates, we deemed it necessary to assign a distinctive name to the high-level products. This not only reflects their enhanced functionality but also communicates to users that these products differ significantly from earlier versions.}
The \pjfinal{UVIT Data Release version 7 products will eventually be provided for all existing observations and are already available in the ISSDC \textit{AstroSat} Archive for a few observations}.
\pjfinal{The UVIT Data Release version 7} products mainly consist of combined UV data products in FITS format. 
The combined UV data products for a single channel-filter-window combination consist of an events list, count-rate and error images, and an exposure map. 
The images and maps (4800$\times$4800 pixels) are provided in both instrument and astronomical coordinate systems.
Products in both systems have astrometry. 
The instrument coordinate system products (also called spacecraft or detector system) will have the orientation of the reference episode data as it was read in the detector---the RA and Dec axes may not be aligned to the image X and Y axes. Meanwhile, the astronomical system products (also called sky or RA-Dec system) will have the RA and Dec axes aligned to the image X and Y axes.
\pjfirst{In the astronomical system products, celestial north is oriented upwards, while east is directed towards the left.}
While the instrument system products have uses in special cases, the astronomical system products are recommended for most users of UVIT data.
The combined VIS quick-look images (1800$\times$1800 pixels) are provided in instrument and astronomical coordinate systems.
While some corrections (distortion and dark) have been applied to the VIS images, no photometric calibration is available for the VIS channel. 
Minor artefacts are introduced in the processing of the VIS channel images that appear as dark spots near sources.
    

\section{Summary}
\textit{AstroSat} UVIT continues to observe the sky even after almost a decade of operations. 
We describe a significant upgrade to the UVIT Level2 pipeline and the newly implemented features.
The high-level data products generated using this new 7.0.1 version of the UVIT pipeline have been posted to ISSDC since June 1, 2024.
The new data products overcome the limitations of the earlier high-level products generated with the previous 6.3 version of the UVIT pipeline.
The latest version of the UVIT pipeline can (i) combine orbit-wise images properly without data loss, and (ii) the astrometry in the combined science-ready final images is accurate to $<$1 arcsecond in most cases (in a few cases where the default astrometry method fails, an alternate astrometry method will provide an accuracy of $<$5 arcseconds).

\section*{Acknowledgements}
\pjfirst{We thank the annonymous reviewers for their suggestions to improve the article.}
Aafitrans \citep{aafitrans}, Astroalign \citep{beroiz2020astroalign}, Astrometry.net \citep{lang2010astrometry}, Astropy \citep{astropy:2013, astropy:2018, astropy:2022}, Curvit \citep{joseph2021curvit}, Joblib, JupyterLab \citep{kluyver2016jupyter}, Matplotlib \citep{hunter2007matplotlib}, NumPy \citep{harris2020array}, Pandas \citep{reback2020pandas, mckinney-proc-scipy-2010}, Photutils \citep{photutils}, Reproject \citep{reproject}, Scipy \citep{2020SciPy-NMeth}, Scikit-image \citep{scikit-image}, SAOImage DS9 \citep{joye2003new}, and TOPCAT \citep{taylor2005topcat} were used in this work. 
The UVIT project is part of the \textit{AstroSat} mission by the Indian Space Research Organisation (ISRO) and includes collaboration from the Indian Institute of Astrophysics (IIA) Bengaluru, The Indian-University Centre for Astronomy and Astrophysics (IUCAA) Pune, Tata Institute of Fundamental Research (TIFR) Mumbai, many centers of ISRO, and the Canadian Space Agency (CSA). This publication uses the data from the \textit{AstroSat} mission of the Indian Space Research Organisation (ISRO), archived at the Indian Space Science Data Centre (ISSDC).
\pjfirst{This work has made use of data from the European Space Agency (ESA) mission
{\it Gaia} (\url{https://www.cosmos.esa.int/gaia}), processed by the {\it Gaia}
Data Processing and Analysis Consortium (DPAC,
\url{https://www.cosmos.esa.int/web/gaia/dpac/consortium}). Funding for the DPAC
has been provided by national institutions, in particular the institutions
participating in the {\it Gaia} Multilateral Agreement.}
\vspace{-1em}


\appendix

\section{Description of the bug in UVIT Level2 pipeline 6.3 version}
\label{sec:bug}

\begin{figure}
\centering
        \includegraphics[width=\columnwidth]{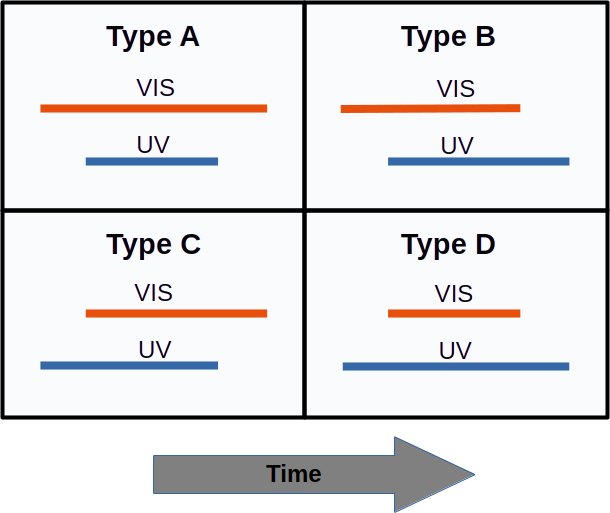}
    \caption{Illustration of the four possible types of temporal overlaps between VIS and UV (FUV/NUV) channel observations in episodes. 
    The horizontal bars represent the duration of VIS (orange) and UV (blue) Level1 data in a single episode.
    }
    \label{fig:four_overlaps}
\end{figure}

\pjfirst{This section outlines a bug identified in UP6 that is corrected in UP7.
The bug manifests under specific conditions and impacts the Level2 data generated with UP6.
The known effects of the bug are:
\begin{itemize}
    \item inaccurate flat-field correction,
    \item incorrect bad-event flagging, and
    \item non-linear jumps in two events list time columns. The two affected columns are \texttt{UVIT\_MASTER\_TIME} and \texttt{MJD\_L2}.
\end{itemize}
The issue appears to be limited to certain episodes where there is partial temporal overlap between VIS and UV Level1 data. 
However, not all cases of partial VIS-UV overlap are affected.
Figure \ref{fig:four_overlaps} illustrates the four possible types of VIS-UV temporal overlaps in an episode:
\begin{itemize} 
\item Type A: the VIS observation starts before and ends after the UV observation.
\item Type B: the VIS observation starts before and ends before the \pjfinal{UV} observation.
\item Type C: the VIS observation starts after the UV observation begins and ends after the UV observation stops.
\item Type D: the VIS observation starts after and ends before the UV observation.
\end{itemize}
It has been observed that only episodes with Type C and Type D overlaps are affected by the bug.
\pjnew{UVIT operation sequence is always planned to conform to the Type A overlap.
However, Level1 data corresponding to the other overlap types (B, C, and D) may arise due to data drops during telemetry transmission and errors during the generation of Level1 data.}
Consequently, the number of episodes impacted by this bug is expected to be minimal.
}

\subsection{\pjnew{Effect of the bug}}

\pjfirst{The flat-field values and bad-event flags for each event are recorded as columns in the Level2 events list.
Our checks reveal that, although UP6 correctly calculates the flat-field values for each event location during initial processing stages, these values become mismatched in the final event list. 
As a result, the flat-field values are associated with incorrect events. The same issue occurs with the bad-event flag column, leading to each event being assigned an incorrect flat-field value and a mismatched bad-event flag. 
\pjmore{For the affected episodes, our calculations indicate that erroneous flat-field corrections result in uncertainties of less than 1\% for \pjmoreii{sources within a radius of 1900 sub-pixels from the centre in} the NUV filter images, except for the N219M and N279N filters. For the N279N filter and all FUV filters, the uncertainties remain below 5\%, while the N219M filter exhibits a maximum uncertainty of less than 10\% \citep{tandon2020additional}.}}
\pjmoreii{All reported uncertainties represent one standard deviation.}

\pjfirst{Additionally, hot pixels that are flagged and masked in the affected episode using UP6 will still appear in the final image due to the mix-up in the bad-event flag column.
The non-linear jumps in the time columns also present a significant issue for users attempting to analyse source variability within the affected episodes.}

\subsection{Identification of episodes affected by the bug}

\pjfirst{Episodes affected by the bug can be identified using the following methods:
\begin{enumerate} 
    \item Temporal overlap check: Analyse the temporal overlaps between VIS and FUV observations within the episodes. Specifically, identify episodes exhibiting Type C and Type D overlaps as described in Fig. \ref{fig:four_overlaps}.
    \item Time consistency check: Examine the \texttt{UVIT\_MASTER\_TIME} and \texttt{MJD\_L2} columns for non-linear jumps by comparing them with the \texttt{FrameCount} column. Under normal conditions, a linear relationship is expected between \texttt{FrameCount} and both \texttt{UVIT\_MASTER\_TIME} and \texttt{MJD\_L2}. 
\end{enumerate}}

\bibliography{references}

\end{document}